\def\ba{\begin{array}}
\def\ea{\end{array}}

\def\dps{\displaystyle}
\def\be{\begin{equation}}
\def\ee{\end{equation}}

\documentclass[aps,pra,showpacs,twocolumn]{revtex4}
\usepackage{amsmath}
\usepackage{amsfonts}
\usepackage{amssymb}

\def\qed{\leavevmode\unskip\penalty9999 \hbox{}\nobreak\hfill
     \quad\hbox{\leavevmode  \hbox to.77778em{%
               \hfil\vrule   \vbox to.675em%
               {\hrule width.6em\vfil\hrule}\vrule\hfil}}
     \par\vskip3pt}

\newcommand{\tr}{\operatorname{Tr}}

\newcommand{\ip}[2]{\left\langle #1 , #2\right\rangle}

\newcommand{\setft}[1]{\mathrm{#1}}
\newcommand{\herm}[1]{\setft{Herm}\left(#1\right)}
\newcommand{\pos}[1]{\setft{Pos}\left(#1\right)}
\newcommand{\ppt}[1]{\setft{PPT}\left(#1\right)}

\begin{document}
\title{$d$ locally indistinguishable maximally entangled states in $\mathbb{C}^d\otimes\mathbb{C}^d$ }
\author{Mao-Sheng Li$^{1}$, Yan-Ling Wang$^{1}$, Shao-Ming Fei$^{2,3}$, Zhu-Jun Zheng$^{1}$}

 \affiliation
 {
 {\footnotesize  {$^1$Department of Mathematics,
 South China University of Technology, Guangzhou
510640, China}} \\
{\footnotesize{
  $^2$School of Mathematical Sciences, Capital Normal University,
Beijing 100048, China}}\\
{\footnotesize{$^3$Max-Planck-Institute for Mathematics in the Sciences, 04103
Leipzig, Germany}}
}

\begin{abstract}
 We give a explicit construction of $d$ locally indistinguishable orthogonal maximally entangled states in
 $\mathbb{C}^d\otimes\mathbb{C}^d$ for any $d\geq 4$.  This gives an answer to the conjecture proposed by
 S. Bandyopadhyay in 2009. Thus it reflects the nonlocality of the fundamental feature of quantum mechanics.
\end{abstract}

\maketitle

In compound quantum systems, global operators can not be implemented generally by using only local operations and classical communication (LOCC).
This reflects the fundamental feature of nonlocality in quantum mechanics. The understanding of the limitation of quantum operators that can be
implemented by LOCC is one of the most important problems in quantum information theory. The local distinguishability of quantum states plays
important roles in exploring quantum nonlocality \cite{Walgate02,Bennett99}. In the bipartite case, Alice and Bob share a quantum system
which is chosen from one of a known set of mutually orthogonal quantum states. Their goal is to identify the given state by using only LOCC.
The nonlocality of quantum information is therefore revealed when a set of orthogonal states can not be distinguished by LOCC.
The local distinguishability has also practical applications in quantum cryptography primitives such as secret sharing and data hiding \cite{Markham08,DiVincenzo02}.

The local distinguishability problem of orthogonal quantum states has received considerable attention in recent years.
Walgate et.al. showed that any two orthogonal pure states can be distinguishable by LOCC \cite{Walgate00}.
In \cite{Nathanson05}, it has been showed that in $\mathbb{C}^3\otimes\mathbb{C}^3$, any three mutually orthogonal maximally
entangled states can be distinguishable by LOCC. It has been observed in \cite{Fan04,Ghosh01,Ghosh04,Nathanson05} that no more than
$d$ maximally entangled states in $\mathbb{C}^d\otimes\mathbb{C}^d$ can be perfectly distinguished.
Since then it has been an interesting open problem if there exit any $N\leq d$ orthogonal maximally entangled states
which are indistinguishable under LOCC \cite{Ghosh04}. Bandyopadhyay conjectured the existence of $d$ or $d-1$ indistinguishable
LOCC maximally entangled states by presenting some sets of quantum states which are one-way LOCC indistinguishable \cite{Ghosh11}.

Since it is difficult to formulate the LOCC in general, one uses the partial-positive transpose (PPT) measurements instead \cite{Nathanson2013,Cosentino13,CosentinoR14,Duan11,Duan14}.
If a set of quantum states can not be distinguished by PPT measurements, then neither can it be distinguished by LOCC.
In \cite{Duan11}, Yu presented four maximally entangled states which are PPT indistinguishable. More recently,
Ref.\cite{Cosentino13} gave a construction of $d=2^n$ PPT-indistinguishable states in
$\mathbb{C}^d\otimes\mathbb{C}^d$. In \cite{CosentinoR14} the authors gave $N<d$ PPT-indistinguishable states in $\mathbb{C}^d\otimes\mathbb{C}^d$ for $d=2^n$, $n>3$.

In this paper, we give an explicit construction of $d$ locally indistinguishable maximally entangled states in $\mathbb{C}^d\otimes\mathbb{C}^d$ for any $d\geq 4$.
This gives an answer to the conjecture proposed by S. Bandyopadhyay in \cite{Ghosh11}. We first use the method in \cite{Cosentino13} to transfer the PPT-distinguishable
problem to a semidefinite program problem. The major difficulty in solving the semidefinite program problem is to find a feasible solution of its dual program.
We show that if the partial transposed operators of given states have a common eigenvector corresponding to negative
eigenvalue, then a feasible solution can be obtained. Moreover, we give a sufficient and necessary condition on feasible solutions of the semidefinite program.
At last, we investigate the case of $d\geq 4$ by
constructing a set of $n^2$ states such that any $2n$ states chosen from this set is PPT-indistinguishable.

Let $A$ and $B$ be the $d$-dimensional complex vector spaces associated with the Alice and Bob's systems.
Let $\herm{A\otimes B}$ and $\pos{A\otimes B}$ denote the sets of all Hermitian operators and positive semidefinite operators on $A\otimes B$ respectively.
We say that $M_1\geq M_2$ if $M_1-M_2$ is positive semidefinite for any Hermitian operators $M_1$ and $M_2$.
Denote $L(A,B)$ the set of all linear maps from $A$ to $B$ ($L(A,A)=L(A)$ for short).
Let $T_A$ be the partial transpose map $T\otimes I_B$ from $A\otimes B$ to $A\otimes B$, where $T$ is the transpose map from $A$ to $A$,
$I_B$ is the identity operator on $B$. We call a positive semidefinite operator $M\in A\otimes B$ a PPT operator
if $T_A(M)\geq 0$. By $\text{PPT}(A:B)$ we denote the set of all PPT operators on the tensor product space $A\otimes B$.

Consider $d$ orthogonal maximally entangled states $\{|\psi_i\rangle\}_{i=1}^{d}$ in $\mathbb{C}^d \otimes \mathbb{C}^d$.
Generally, $|\psi_i\rangle=(I\otimes U_i) |\psi_1\rangle$, where $|\psi_1\rangle=\frac{1}{\sqrt{d}}\sum_{i=1}^{d}|ii\rangle$,
and $U_i$  are unitary matrices. Since there is a one to one correspondence between a maximally entangled state $|\psi_i\rangle$
and the unitary matrix $U_i$, we call the unitary matrices $\{U_i\}_{i=1}^{d}$ the defining unitary matrices of the maximally
entangled states $\{|\psi_i\rangle\}_{i=1}^{d}$.
With respect to the pure state $|\psi_i\rangle$, the corresponding density matrix is given by
\be\label{rhoi}
\rho_{i}=|\psi_i\rangle\langle\psi_i|.
\ee
A set of states $\{|\psi_i\rangle\}_{i=1}^d$ is called PPT-distinguishable if there exist PPT measurements $\{P_i\}_{i=1}^d$
such that $\ip{P_i}{\rho_{j}}=\delta_{ij}$, namely, $ \frac{1}{d}\sum_{j = 1}^d  \ip{P_j}{\rho_{j}}=1$.
Otherwise, the set $\{|\psi_i\rangle\}_{i=1}^d$ is said to be PPT-indistinguishable.
To find the maximal success probability of distinguishing the set of states $\{|\psi_i\rangle\}_{i=1}^{d}$ with PPT measurements is
equivalent to the following semidefinite program \cite{Cosentino13},
\be\label{sdp-primal}
\text{maximize}~ \frac{1}{d}\sum_{j = 1}^d  \ip{P_j}{\rho_{j}},
\ee
subject to $P_1+ \cdots + P_d = I_{A} \otimes I_{B}$, and $P_1,\ldots,P_d\in\ppt{A:B}$.
The dual problem \cite{Cosentino13} to minimize $\tr(\gamma)/d$,
subject to $\gamma - \rho_{j} \geq T_{A}(Q_{j})$ for $j=1,\ldots,d$, $\gamma\in\herm{A\otimes B}$, $Q_{1}, \ldots, Q_{d}\in\pos{A\otimes B}$,
which is still difficult to tackle with. If one further constrains the dual problem by imposing equality instead of
inequality constraints in the above program, one gets the following program:
\be\label{sdp-dual-moreconstrained}
\text{minimize}~ \frac{1}{d}\tr(\gamma),
\ee
subject to $\gamma \geq T_{A}(\rho_{j})$ for $j=1,\ldots,d$, $\gamma\in\herm{A\otimes B}$.
As has been pointed out in \cite{Cosentino13}, any feasible solution of program (\ref{sdp-dual-moreconstrained}) provides an upper bound
of program (\ref{sdp-primal}). In the following we study the program (\ref{sdp-dual-moreconstrained}) and give a
sufficient and necessary condition for program (\ref{sdp-dual-moreconstrained}) to have a feasible solution.

\noindent {\bf Theorem 1.} For any $d\geq 4$, there exist $d$ PPT-indistinguishable, hence LOCC indistinguishable,
mutually orthogonal maximally entangled states in $\mathbb{C}^d \otimes \mathbb{C}^d$.

These $d$ PPT-indistinguishable mutually orthogonal maximally entangled states in $\mathbb{C}^d \otimes \mathbb{C}^d$ have to be constructed
separately for $d=4n$ ($n\geq 1$), $d=4n+1$, $d=4n+3$ ($n\geq 2$) and $d=5,7,11$. Before proving theorem 1, let us first
introduce some useful results. For given $\rho_{i}=|\psi_i\rangle\langle\psi_i|$, $|\psi_i\rangle=(I\otimes U_i) |\psi_1\rangle$,
one has $T_{A}(\rho_{i})=(I\otimes U_i)T_A(\rho_1)(I\otimes U_i^{\dagger})$. In particular,
the eigenvalues of $T_A(\rho_1)$ are either $\frac{1}{d}$ or $-\frac{1}{d}$.

\noindent {\bf Lemma} Let $V_{\lambda}(M)$ denote the set of all eigenvectors of an $n\times n$ matrix $M$ corresponding to an eigenvalue $\lambda$. Then
$$
\ba{l}
V_{-\frac{1}{d}}(T_A(\rho_1))=\text{span}\{|kl\rangle-|lk\rangle,1\leq k<l\leq d\},\\[1mm]
V_{\frac{1}{d}}(T_A(\rho_1))=\text{span}\{|kl\rangle+|lk\rangle,1\leq k\leq l\leq d\}.
\ea
$$
The number of linear independent eigenvectors of $T_A(\rho_1))$ corresponding to the eigenvalue $-\frac{1}{d}$ and $\frac{1}{d}$ are
$L=\frac{(d-1)d}{2}$ and $M=\frac{(d+1)d}{2}$, respectively.

\noindent {\bf Theorem  2.} For the states $\rho_i$ defined in (\ref{rhoi}), we have
$\bigcap_{i=1}^{d}  V_{-\frac{1}{d}}(T_A(\rho_i))\neq\{0\}$
if and only if  there is a feasible solution of the semidefinite program (3) satisfying
$$\gamma\leq \frac{1}{d}I_A\otimes I_B, \text{ with } \gamma\neq \frac{1}{d}I_A\otimes I_B.$$

\noindent \emph{Proof:} Suppose $0\neq|v\rangle \in \bigcap_{i=1}^{d} V_{-\frac{1}{d}}(T_A(\rho_i))$ for $i=1,...,d$,
that is, $ T_A(\rho_i) |v\rangle=-\frac{1}{d}|v\rangle$ for $i=1,...,d$. As $T_{A}(\rho_{i})$ is an Hermitian unitary matrix,
from singular value decomposition there exist orthogonal normal vectors $|v_i^l\rangle$ and $|w_i^m\rangle$ such that
$$
\ba{l}
\dps T_{A}(\rho_{i})=-\frac{1}{d}|v\rangle\langle v|-\frac{1}{d}\sum_{l=2}^L|v_i^l\rangle\langle v_i^l|+\frac{1}{d}\sum_{m=1}^M|w_i^m\rangle\langle w_i^m|,\\[2mm]
I_A\otimes I_B=|v\rangle\langle v|+\displaystyle\sum_{l=2}^L|v_i^l\rangle\langle v_i^l|+\sum_{m=1}^M|w_i^m\rangle\langle w_i^m|.
\ea
$$
Hence we have
$$
\frac{1}{d}I_A\otimes I_B-\frac{2}{d}|v\rangle\langle v|-T_{A}(\rho_{i})=\frac{2}{d}\sum_{l=2}^L|v_i^l\rangle\langle v_i^l|\geq0.
$$
Clearly, $\gamma=\frac{1}{d}I_A\otimes I_B-\frac{2}{d}|v\rangle\langle v|\in \herm{A\otimes B}$. Therefore $\gamma $ is a feasible
solution of semidefinite program (\ref{sdp-dual-moreconstrained}) which satisfies $ \gamma\leq \frac{1}{d}I_A\otimes I_B$ and $ \gamma\neq \frac{1}{d}I_A\otimes I_B$ .

Conversely, if $\gamma$ is a feasible  solution satisfying $\gamma\leq \frac{1}{d}I_A\otimes I_B,$ we can suppose
$\frac{1}{d}I_A\otimes I_B-\gamma=\sum_{k=1}^K\mu_k|v^k\rangle\langle v^k|$ with $\mu_k\geq0$ for all $k=1,...,K$ and at
least one of the $\mu_k$ strictly positive, say, $\mu_1>0$. Since $\gamma$ is a feasible solution,
we have the following inequalities for all $i=1,...,d$,
$$
\frac{1}{d}I_A\otimes I_B-\sum_{k=1}^K\mu_k|v^k\rangle\langle v^k|\geq T_{A}(\rho_{i}),
$$
which implies
\begin{equation}\label{4}
\frac{1}{d}I_A\otimes I_B-T_{A}(\rho_{i})\geq \mu_1|v^1\rangle\langle v^1|.
\end{equation}

For each $1\leq i\leq d$, $T_{A}(\rho_{i})$ has the following singular value decomposition,
\begin{equation}\label{5}
T_{A}(\rho_{i})=-\frac{1}{d}\sum_{l=1}^L|v_i^l\rangle\langle v_i^l|+\frac{1}{d}\sum_{m=1}^M|w_i^m\rangle\langle w_i^m|,
\end{equation}
where $\{|v_i^l\rangle\}_{l=1}^L\cup \{|w_i^m\rangle\}_{m=1}^M$ form an orthogonal normal base of  $\mathbb{C}^d \otimes \mathbb{C}^d.$
Hence we have the following identity,
\begin{equation}\label{6}
I_A\otimes I_B=\sum_{l=1}^L|v_i^l\rangle\langle v_i^l|+\sum_{m=1}^M|w_i^m\rangle\langle w_i^m|.
\end{equation}
From (\ref{4}), (\ref{5}) and (\ref{6}) we have
$$
\frac{2}{d}\sum_{l=1}^L|v_i^l\rangle\langle v_i^l|\geq \mu_1|v^1\rangle\langle v^1|.
$$
Therefore  $|v^1\rangle \in \text{span} \{|v_i^l\rangle\}_{l=1}^L.$
From the singular value decomposition (\ref{5}), one has that $\text{span} \{|v_i^l\rangle\}_{l=1}^L$ is just the set of the
eigenvectors of $T_{A}(\rho_{i})$ corresponding to the eigenvalue $-\frac{1}{d}$. Hence $|v^1\rangle$ must be an eigenvector of $T_{A}(\rho_{i})$ corresponding to the eigenvalue $-\frac{1}{d}$.
That is, $T_{A}(\rho_{i})|v^1\rangle=-\frac{1}{d}|v^1\rangle$. Hence $ |v^1\rangle \in \bigcap_{i=1}^d  V_{-\frac{1}{d}}(T_A(\rho_i))$. \qed

\noindent {\bf Theorem  3.}  $ \bigcap_{i=1}^{d}  V_{-\frac{1}{d}}(T_A(\rho_i))\neq\{0\} $ if and only if
 $ \bigcap_{i=1}^{d} (I\otimes U_i^\dag) V_{-\frac{1}{d}}(T_A(\rho_1))\neq\{0\}. $ \\

\emph{Proof:} If $0\neq|v\rangle \in \bigcap_{i=1}^{d}  V_{-\frac{1}{d}}(T_A(\rho_i))\neq\{0\}$, that is,
$T_A(\rho_i)|v\rangle=-\frac{1}{d}|v\rangle$, $i=1,2,...,d.$ We have
$(I\otimes U_i)T_A(\rho_1)(I\otimes U_i^{\dagger})|v\rangle=-\frac{1}{d}|v\rangle$,
and $T_A(\rho_1)(I\otimes U_i^{\dagger})|v\rangle=-\frac{1}{d}(I\otimes U_i^{\dagger})|v\rangle$.
Hence $|v\rangle\in \bigcap_{i=1}^{d} (I\otimes U_i^\dag) V_{-\frac{1}{d}}(T_A(\rho_1))$.
The converse can be proved straightforwardly.\qed

\noindent {\bf Corollary} If $ \bigcap_{i=1}^{d} (I\otimes U_i^\dag) V_{-\frac{1}{d}}(T_A(\rho_1))\neq\{0\},$
then the set $\{ |\psi_i\rangle \}_{i=1}^{d}$ defined by $\{ U_i \}_{i=1}^{d}$ is PPT indistinguishable.
Particularly, $\{ |\psi_i\rangle \}_{i=1}^{d}$ is LOCC indistinguishable.

\noindent {\it Remark} If $|v\rangle \in \bigcap_{i=1}^{d} (I\otimes U_i^\dag) V_{-\frac{1}{d}}(T_A(\rho_1)),$  we
have $|v\rangle \in V_{-\frac{1}{d}}(T_A(\rho_1))$ as $U_1=I_B.$ Hence the matrices $I\otimes U_i^\dag$
transform the same $|v\rangle \in V_{-\frac{1}{d}}(T_A(\rho_1))$ to the eigenvectors of
$T_A(\rho_1)$ with eigenvalue $-\frac{1}{d}$.

We now prove the theorem 1 by investigating the following cases:

\noindent\textbf{\emph{Case I: d=2n.}}
Set $w=e^\frac{2\pi \sqrt{-1}}{n}$. We construct the $2n$ orthogonal unitary matrices as follows:
$$
\ba{rcl}
U_1&=&diag(1,w^0,...,w^{0(n-1)},1,w^0,...,w^{0(n-1)}),\\[1mm]
U_2&=&diag(1,w,...,w^{n-1},1,w,...,w^{n-1}),\\[1mm]
&\cdots&\\[1mm]
U_n&=&diag(1,w^{n-1},...,w^{{(n-1)}^2},1,w^{n-1},...,w^{{(n-1)}^2}),\\[1mm]
U_{n+2}&=&diag(1,w,...,w^{{n-1}},w^{{n-1}},1,w,...,w^{{n-2}})U_{n+1},\\[1mm]
&\cdots&\\[1mm]
U_{2n}&=&diag(1,w^{{(n-1)}\times1},...,w^{{(n-1)}^2},w^{{(n-1)}^2},1,\\[1mm]
&&w^{{(n-1)}\times 1},...,w^{(n-1)(n-2)})U_{n+1},
\ea
$$
while
$$
U_{n+1}=\left[
\begin{array}{cc}
S &  \textbf{0}   \\
   \textbf{0} & S^T\end{array}\right],
$$
where
$$
S^T=\left[
                 \begin{array}{ccccc}
                   0 & 1 & 0 & \cdots & 0 \\
                   0 & 0 & 1 & \cdots & 0 \\
                   \vdots & \vdots & \vdots & \ddots & \vdots \\
                   0 & 0 & 0 & \cdots & 1 \\
                   1 & 0 & 0 & \cdots & 0 \\
                 \end{array}\right].
$$

In fact, the above unitary matrices can be defined as the first $2n$ matrices of the following $n^2$ orthogonal
unitary matrices $\{U_{kn+l}\}$ which, under the computational base $\{|m\rangle\}_{m=1}^{2n}$, can be expressed as
$$
\ba{l}
\dps\sum_{m=1}^n (w^{(m+k-1)(l-1)}|k\oplus m\rangle\langle m|\\[3mm]
\dps~~~~~~~~~~~+ w^{(m-k-1)(l-1)}|n+m\rangle\langle n+(k\oplus m)|),
\ea
$$
where $k=0,...,n-1$, $l=1,...,n$ and $k\oplus m$ stands for the number $k+m \text{ mod n }$.
As
$$
\ba{l}
\dps(I\otimes U_{kn+l}) \sum_{m=1}^{n}(|m\rangle|n+m\rangle-|n+m\rangle|m\rangle)\\[2mm]
~~~=\dps\sum_{m=1}^{n}w^{(m+k-1)(l-1)}(|k\oplus m\rangle|n+m\rangle\\[5mm]
~~~~~~~\dps -|n+m\rangle |k\oplus m\rangle).
\ea
$$
From Lemma we notice that the right hand side of the above equality is also in $V_{-\frac{1}{d}}(T_A(\rho_1))$.
Set
$$
|v\rangle=\sum_{m=1}^{n}(|m\rangle|n+m\rangle-|n+m\rangle|m\rangle).
$$
Then
\begin{equation}
 |v\rangle\in \bigcap_{i=1}^{n^2}( I\otimes U_i) V_{-\frac{1}{d}}(T_A(\rho_1)).
 \end{equation}
In particular,  we obtain
$$
|v\rangle\in \bigcap_{i=1}^{d} (I\otimes U_i) V_{-\frac{1}{d}}(T_A(\rho_1)).
$$
Therefore we can conclude that
the $2n$ states  $\{ |\psi_i\rangle=(I\otimes U_i^\dag) |\psi_1\rangle\}_{i=1}^{2n}$ are PPT-indistinguishable.

\noindent\textbf{\emph{ Case II: d=2n+1.}}
Now we give a construction of $2n+1$ PPT-indistinguishable states.
We deal with the problem according to (i) $d=4n+1$, $n\geq 2$ and (ii) $d=4n+3$, $n\geq 3$.
The cases for $n=5,7,11$ will be considered separately.

(i). We construct $U_j$ to be block unitary matrices of the form $diag(V_j, W_j)$,
where $V_j$ are $(2n+2)\times (2n+2)$ matrices and  $W_j$ are $(2n-1)\times (2n-1)$ matrices.
We chose $V_j$ to be the $(n+1)^2$ matrices defined above in the \textbf{\emph{Case I}} with $d=2(n+1)$, and
$W_j$ the $(2n-1)^2$ generalized $(2n-1)\times (2n-1)$ Pauli matrices defined by
$\{ X^aZ^b| a,b=0,1,...,n-1\}$, where
$X=|0\rangle\langle n-1|+\sum_{i=0}^{n-2}|i+1\rangle\langle i|,$
$Z=\sum_{i=0}^{n-1}w^i|i\rangle\langle i|$, $w=e^\frac{2\pi \sqrt{-1}}{n} $, and $\{|i\rangle\}_{i=0}^{n-1}$ is the computational basis.
If $n\geq2$, we have $(n+1)^2\geq 4n+1$ and $(2n-1)^2\geq 4n+1.$  So we can really construct $4n+1$ orthogonal unitary matrices $\{U_j\}$.

(ii). Similar to the above construction, we assume $U_j$ be of the form $diag(V_j, W_j)$,
where $V_j$ are $(2n+2)\times (2n+2)$ matrices and $W_j$ are $(2n+1)\times (2n+1)$ matrices.
Suppose $V_j$ are chosen from the $(n+1)^2$ matrices defined above in the \textbf{\emph{Case I}} for $d=2(n+1)$.
And $W_j$ are chosen from the $(2n+1)^2$ generalized $(2n+1)\times (2n+1)$ Pauli matrices.
If $n\geq 3$, we have $(n+1)^2\geq 4n+3$ and $(2n+1)^2\geq 4n+3$. Then we can construct $4n+3$ orthogonal unitary matrices $\{U_j\}$.

In these two cases, the orthogonality of $\{U_j\}$ can be derived from the orthogonality of $\{V_j\}$ and the generalized Pauli matrices $\{V_j\}.$
Denote $|v\rangle=\sum_{k=1}^{n+1}(|k\rangle|n+1+k\rangle-|n+1+k\rangle|k\rangle)$. It is easily verified that
$$
|v\rangle\in \bigcap_{i=1}^{d} (I\otimes U_i)V_{-\frac{1}{d}}(T_A(\rho_1)).
$$
Hence we can conclude that when $n\geq2$, $d=4n+1$ or $n\geq3$, $d=4n+3$, the $d$ states
$\{ |\psi_i\rangle=(I\otimes U_i^\dag) |\psi_1\rangle\}_{i=1}^{2n+1}$ are PPT-indistinguishable.

\noindent\textbf{\emph{ Case III: d=5,7,11.}}
We consider now the exceptional cases of odd $d$, $d=5,7,11$.
For $d=5$, $5$ PPT indistinguishable states  have been presented in \cite{Cosentino13}.
Foe $d=7$, we chose $U_i=diag(V_i, W_i)$,
where 

$$
  V_1=diag(1, 1, 1, 1), ~~~~ V_6=diag(1, -1, 1, -1),
$$ 
$$
  V_2=\left[\begin{array}{cccc}
         &   &   &1   \\
         &   & 1  &   \\
         & 1  &   &   \\
        1 &   &   &
     \end{array}
     \right],~~~~
  V_3=\left[\begin{array}{cccc}
         &   &   &1   \\
         &   & 1  &   \\
         &  -1 &   &   \\
         -1&   &   &
     \end{array}
     \right],
$$
$$
  V_4=\left[\begin{array}{cccc}
         &   1&   &   \\
         1&   &   &   \\
         &   &   &-1   \\
         &   & -1  &
     \end{array}
     \right].~~~~
 V_5=\left[
             \begin{array}{cccc}
                & 1 &  &  \\
                1 &  &  &  \\
                &  &  & 1  \\
                &  & 1 &  \\
             \end{array}
           \right],
$$
$W_{i+1}=diag(1,\omega^i,\omega^{2i})$, $W_{i+4}=diag(1,\omega^i,\omega^{2i})\,S$ for $i=0,1,2$, and
$$
S=\left[
           \begin{array}{ccc}
             0 & 0 & 1 \\
             1 & 0 & 0 \\
             0 & 1 & 0 \\
           \end{array}
         \right].
$$

Set $|v\rangle=\frac{1}{2}(|13\rangle-|31\rangle+|24\rangle-|42\rangle)$. We have
$$
|v\rangle\in \bigcap_{i=1}^{6}(I\otimes U_i^\dag)V_{-\frac{1}{7}}(\rho_1).
$$
Let $\gamma=\frac{1}{7}I_A\otimes I_B-\frac{2}{7}|v\rangle\langle v|$.
Then $\gamma\geq T_{A}(\rho_{i})$, where $\rho_{i}$ is given by $|\psi_i\rangle=(I\otimes U_i) |\psi_1\rangle$, $i=1,2,\cdots,6.$
We chose the seventh unitary matrix $U_7$ to be
$$
U_7=\left[
            \begin{array}{ccccccc}
               0& 0 & 0 & 0 & 0 & 1 & 0 \\
              0  & 0 & 0 & 0 & 0 & 0 & 1 \\
               1& 0 & 0 & 0 & 0 & 0 & 0 \\
               0& 1 & 0 & 0 & 0  & 0 & 0 \\
               0&  0& 1 & 0 & 0 & 0 & 0 \\
               0&  0& 0 & 1 & 0 & 0 & 0 \\
              0 & 0 & 0 & 0 & 1 & 0 & 0 \\
            \end{array}
          \right].
$$
$U_7$ is orthogonal to the above six unitary matrices $\{U_i\}_{i=1}^6$.
From $U_7$ one has $|\psi_7\rangle=(I\otimes U_7) |\psi_1\rangle.$
After a lengthy calculation, we obtain
$|v\rangle=\sqrt{\frac{3}{8}}|u\rangle+\sqrt{\frac{5}{8}}|w\rangle,$
where $|u\rangle$ and $|w\rangle$ are the eigenvectors of $T_A(\rho_{7})$ with
$\rho_{7}=|\psi_7\rangle\langle \psi_7|$,
$T_A(\rho_{7})|u\rangle=-\frac{1}{7}|u\rangle$ and
$T_A(\rho_{7})|w\rangle=\frac{1}{7}|w\rangle $, satisfying $\langle u|u\rangle=\langle w|w\rangle=1$.

Unfortunately, we can not find a feasible solution of programm (\ref{sdp-dual-moreconstrained}) with
the form given by Theorem $2$. Instead, we find a feasible solution of program (\ref{sdp-dual-moreconstrained})
with the form
$$
\widetilde{\gamma}=\frac{1}{7}_A\otimes I_B-\frac{\lambda}{7}|v\rangle\langle v|+\frac{\mu}{7}|w\rangle\langle w|.
$$
Then we must have
\begin{equation}\label{8}
\frac{1}{7}I_A\otimes I_B-\frac{\lambda}{7}|v\rangle\langle v|+\frac{\mu}{7}|w\rangle\langle w|>T_{A}(\rho_{7}).
\end{equation}
Suppose that the singular value decomposition of $T_{A}(\rho_{7})$ have the form:
$$
T_{A}(\rho_{7})=\frac{1}{7}(-|u\rangle\langle u|-\sum_{l=2}^L |u_l\rangle\langle u_l|+|w\rangle\langle w|+\sum_{m=2}^M |w_m\rangle\langle w_m|).
$$
We have
$$
I_A\otimes I_B=|u\rangle\langle u|+\sum_{l=2}^L |u_l\rangle\langle u_l|+|w\rangle\langle w|+\sum_{m=2}^M |w_m\rangle\langle w_m|.
$$
A direct calculation shows that the inequality (\ref{8}) is equivalent to
$$
\frac{1-\frac{3}{8}\lambda}{7}|u\rangle\langle u|-\frac{\frac{\sqrt{15}}{8}\lambda}{7}(|u\rangle\langle w|+|w\rangle\langle u|)+\frac{\mu-\frac{5}{8}\lambda}{7}|w\rangle\langle w|\geq 0,
$$
which is satisfied if the following inequalities hold.
$$\left\{
             \begin{array}{ll}
               1-\frac{3}{8}\lambda>0, \\[3mm]
               (1-\frac{3}{8}\lambda)(\mu-\frac{5}{8}\lambda)-\frac{15}{64}\lambda^2>0.
             \end{array}
           \right.
$$
To find a feasible solution $\widetilde{\gamma}$ with $\frac{1}{7}Tr(\widetilde{\gamma})<1$,  must have $\mu<\lambda$. Hence we  get $\mu<\lambda<{\mu}({\frac{5}{8}+\frac{3}{8}\mu})^{-1}.$
By choosing $\mu={1}/{2}$ and $\lambda={7}/{13}$, the corresponding $\widetilde{\gamma}$ satisfies
$\widetilde{\gamma}\geq T_{A}(\rho_7)$, which follows from $\gamma\geq T_{A}(\rho_{i})$ for $i=1,2,\cdots,6$.
Therefore $\widetilde{\gamma}$ is a feasible solution of the program (\ref{sdp-dual-moreconstrained}) with $Tr(\widetilde{\gamma})/7<1$.
Hence the seven states defined above are PPT-indistinguishable.

At last, consider $d=11$. Assume $U_j=diag(V_j,W_j)$ with
$V_j$ $6\times 6$ unitary matrices and $W_j$ $5\times 5$ unitary matrices.
Suppose that the $6\times 6$ unitary matrices $V_j$ are chosen from the following ones,
$$
\ba{rcl}
 V_i&=&(|1\rangle\langle 1|+|2\rangle\langle 2|)+w^i(|3\rangle\langle 3|\\[1mm]
 &&+|4\rangle\langle 4|)+w^{2i}(|5\rangle\langle 5|+|6\rangle\langle 6|),\\[2mm]
 V_{3+i}&=&(|3\rangle\langle 2|-|1\rangle\langle 4|)-w^i(|4\rangle\langle 6|\\[1mm]
 &&+|5\rangle\langle 3|)+w^{2i}(|2\rangle\langle 5|-|6\rangle\langle 1|),\\[2mm]
 V_{6+i}&=&(|1\rangle\langle 3|+|4\rangle\langle 2|)+w^i(|5\rangle\langle 1|\\[1mm]
 &&-|2\rangle\langle 6|)+w^{2i}(|3\rangle\langle 5|+|6\rangle\langle 4|),\\[2mm]
 V_{9+i}&=&(|5\rangle\langle 2|-|1\rangle\langle 5|)-w^i(|2\rangle\langle 4|\\[1mm]
 &&+|3\rangle\langle 1|)+w^{2i}(|4\rangle\langle 6|-|6\rangle\langle 3|),
\ea
$$
where $i=1,2,3.$  If we choose  the $12$ unitary matrices $W_j$ from the generalized  $5 \times 5 $  Pauli matrices, then the twelve unitary matrices $\{U_j\}$ are orthogonal each others. Moreover, $|v\rangle=(|12\rangle-|21\rangle)+(|34\rangle-|43\rangle)+(|56\rangle-|65\rangle)$ satisfies
$$
|v\rangle \in \bigcap_{j=1}^{12} (I\otimes U_j) V_{-\frac{1}{11}}(T_A(\rho_1)).
$$
This implies that the eleven states defined by $\{ |\psi_i\rangle=(I\otimes U_i^\dag) |\psi_1\rangle\}_{i=1}^{11}$ are PPT-indistinguishable.

Summarizing the above results for \textbf{\emph{Case I,II,III}}, we obtain the Theorem $1$. \qed

We have proved the existence of $d$ PPT-indistinguishable
mutually orthogonal maximally entangled states in $\mathbb{C}^d \otimes \mathbb{C}^d$ for any $d\geq 4$, by presenting the detailed constructions
of these states. Such constructions are in fact might be not unique. As an example, in the following we
present a construction for $d=4n$, which is different from the {\emph{ Case I}} even $n=2m$ for some integer $m$.

First, we recall the four PPT indistinguishable states in $\mathbb{C}^4\otimes\mathbb{C}^4$ given in \cite{Duan11}.
The corresponding defining unitary matrices are given in \textbf{\emph{Case III}}.
It is direct to verify  the following identities:
\begin{eqnarray*}
(I\otimes V_2^\dag)(|13\rangle-|31\rangle+|42\rangle-|24\rangle)\\
~~~~~~~~~~~~=|12\rangle-|34\rangle+|43\rangle-|21\rangle,\\
(I\otimes V_3^\dag)(|13\rangle-|31\rangle+|42\rangle-|24\rangle)\\
~~~~~~~~~~~~=|12\rangle+|34\rangle-|43\rangle-|21\rangle,\\
(I\otimes V_4^\dag)(|13\rangle-|31\rangle+|42\rangle-|24\rangle)\\
~~~~~~~~~~~~=-|14\rangle-|32\rangle+|41\rangle+|23\rangle.
\end{eqnarray*}
From Lemma,  we see that all the vectors on the right hand side of the above equalities belong to $V_{-\frac{1}{4}}(T_A(\rho_1)).$
Hence we have
$$
(|13\rangle-|31\rangle+|42\rangle-|24\rangle)\in\bigcap_{i=1}^4 (I\otimes V_i^\dag) V_{-\frac{1}{4}}(T_A(\rho_1)).
$$
By Corollary, we have that the above four states are PPT-indistinguishable, as has been proven in \cite{Cosentino13}.

Based on the above four states, we give a construction of $4n$ PPT-indistinguishable states in  $\mathbb{C}^{4n}\otimes\mathbb{C}^{4n}.$
Let $W_i=diag(1,w^{i},w^{2i},...,w^{(n-1)i})$, $i=0,...,n-1$, where $w=e^\frac{2\pi \sqrt{-1}}{n}.$ We define
$$\ba{ll}
  U_{i+1}=W_{i}\otimes V_1 ,~~~~&U_{n+i+1}=W_{i}\otimes V_2,\\[2mm]
 U_{2n+i+1}=W_{i}\otimes V_3,~~~~&U_{3n+i+1}=W_{i}\otimes V_4
\ea
$$
for $i=0,1,...,n-1$.
Clearly we have
\begin{eqnarray*}
(I\otimes U_{i}^\dag)(|13\rangle-|31\rangle+|42\rangle-|24\rangle)\\
~~~~~~=|13\rangle-|31\rangle+|42\rangle-|24\rangle,\\
(I\otimes U_{n+i}^\dag)(|13\rangle-|31\rangle+|42\rangle-|24\rangle)\\
~~~~~~=|12\rangle-|34\rangle+|43\rangle-|21\rangle,\\
(I\otimes U_{2n+i}^\dag)(|13\rangle-|31\rangle+|42\rangle-|24\rangle)\\
~~~~~~=|12\rangle+|34\rangle-|43\rangle-|21\rangle,\\
(I\otimes U_{3n+i}^\dag)(|13\rangle-|31\rangle+|42\rangle-|24\rangle)\\
~~~~~~=-|14\rangle-|32\rangle+|41\rangle+|23\rangle,
\end{eqnarray*}
for $i=1,2,...,n.$ Therefore we have
$
(|13\rangle-|31\rangle+|42\rangle-|24\rangle)\in \bigcap_{i=1}^{d} (I\otimes U_i) V_{-\frac{1}{d}}(T_A(\rho_1)).
$

Moreover, one can check that $|4k+1,4k+3\rangle-|4k+3,4k+1\rangle+|4k+4,4k+2\rangle-|4k+2,4k+4\rangle\in \bigcap_{i=1}^{d} (I\otimes U_i^\dag) V_{-\frac{1}{d}}(T_A(\rho_1)) $ for any $k=0,1,...,n-1.$ We obtain
$$
\text{dim}(\bigcap_{i=1}^{d} (I\otimes U_i^\dag) V_{-\frac{1}{d}}(T_A(\rho_1)))\geq n.
$$
Then by Corollary, we conclude that the $4n$ states defined by
$\{ |\psi_i\rangle=(I\otimes U_i) |\psi_1\rangle\}_{i=1}^{4n}$ are PPT-indistinguishable.

\medskip

We have studied the locally indistinguishable maximally entangled states in $\mathbb{C}^d\otimes\mathbb{C}^d$ with $d\geq 4$.
From the investigation of the semidifinite program, which is
equivalent to the PPT-indistinguishable problem, we have obtained a sufficient and necessary condition for
its weak dual semidefinite program having a feasible solution of particular form. We have presented an
explicit construction of the $d$ mutually orthogonal maximally entangled states which are
PPT-indistinguishable in $\mathbb{C}^d\otimes\mathbb{C}^d$ for
any $d\geq 4$. Our results also give an answer to the conjecture proposed by S. Bandyopadhyay in \cite{Ghosh11}.

\vspace{2.5ex}
\noindent{\bf Acknowledgments}\, \,
This work is supported by the NSFC 11475178 and NSFC 11275131.


\begin{thebibliography}{}

\bibitem{Bennett99}
C.H. Bennett, D.P. DiVincenzo, C.A. Fuchs, T. Mor, E. Rains, P.W. Shor, J.A. Smolin, and W.K. Wootters,
\newblock {\em Phys. Rev. A}, 59:1070-1091, (1999).

\bibitem{Walgate02}
J. Walgate,  L. Hardy,
\newblock {\em Phys. Rev. Lett}, 89, 147901 (2002).

\bibitem{DiVincenzo02}
D.P. DiVincenzo, D.W. Leung  and B.M. Terhal,
\newblock {\em IEEE Trans. Inf. Theory} 48, 580 (2002).

\bibitem{Markham08}
D. Markham  and B. C. Sanders.
\newblock{ \em Phys. Rev. A}  78, 042309 (2008).

\bibitem{Walgate00}
J. Walgate, A. J. Short, L. Hardy, and V. Vedral,
\newblock {\em Phys. Rev. Lett} 85, 4972 (2000).

\bibitem{Nathanson05}
M. Nathanson,
\newblock {\em J. Math. Phys.} 46, 062103 (2005)

\bibitem{Ghosh01}
S. Ghosh, G. Kar, A. Roy, A. Sen(De), and U. Sen,
\newblock {\em Phys. Rev. Lett} 87, 277902, (2001).

\bibitem{Ghosh04}
S. Ghosh, G. Kar, A. Roy, and D. Sarkar.
\newblock {\em Phys. Rev. A} 70, 022304, (2004).

\bibitem{Fan04}
H. Fan,
\newblock {\em Phys. Rev. Lett} 92, 177905 (2004).

\bibitem{Ghosh11}
S. Bandyopadhyay, S. Ghosh, and G. Kar,
\newblock {\em New J. Phys.} 13 123013, (2011).

\bibitem{Duan11}
N. Yu, R. Duan, and M. Ying,
\newblock {\em Phys. Rev. Lett} 109, 020506 (2012).

\bibitem{Duan14}
N. Yu, R. Duan, and M. Ying,
\newblock {\em IEEE Trans. Inf. Theory }  Vol. 60,  No.4 (2014).

\bibitem{Nathanson2013}
M.~Nathanson,
\newblock {\em Phys. Rev. A}, 88, 062316, (2013).

\bibitem{Cosentino13}
A.~Cosentino,
\newblock {\em Phys. Rev. A}, 87, 012321, (2013).

\bibitem{CosentinoR14}
A.~Cosentino and V.~Russo,
\newblock {\em Quantum Information \& Computation}, 14, 1098--1106, (2014).


\end{thebibliography}
\end{document}